\documentclass[showpacs, twocolumn]{revtex4-1}
\usepackage{graphicx}
\usepackage{amsmath}
\usepackage{appendix}

\begin{document}

\title{ Superfluidity and coherence in uniform dipolar binary Bose mixtures }

\author{Abdel\^{a}ali Boudjem\^{a}a$^{1,2}$}
\affiliation{$^1$ Department of Physics, Faculty of Exact Sciences and Informatics, 
Hassiba Benbouali University of Chlef, P.O. Box 78, 02000, Ouled-Fares, Chlef, Algeria. \\
$^2$Laboratory of Mechanics and energy, Hassiba Benbouali University of Chlef, P.O.Box 78, 02000, Ouled-Fares, Chlef, Algeria.}
\email {a.boudjemaa@univ-chlef.dz}

\date{\today}

\begin{abstract}

We investigate the  superfluidity and the coherence in dipolar binary Bose mixtures using the hydrodynamic approach.
Useful analytical formulas for the excitations spectrum, the correlation function, the static structure factor, and  the superfluid fraction are derived.  
We find that in the case of highly imbalanced mixture, the superfluidity can occur in the dilute component only at extremely low temperatures.
The behavior of the first-order correlation function for both dipolar and nondipolar Bose mixtures is deeply analyzed.
Then we face the two-dimensional case which encodes a non-trivial physics due to the roton modes.

\end{abstract}

\pacs{03.75.Hh, 67.60.Bc, 03.75.Mn, 67.85.Bc} 

\maketitle

\section{Introduction} \label{Intro}

Superfluidity of either fermionic or bosonic ultracold gases is one of the most exciting  phenomena in condensed matter systems.
In the last years, superfluidity in ultracold atomic gas mixtures  have attracted a great interest from both the experimental and theoretical sides
owing of their potential applications in different areas of physics.

Mixtures of superfluid liquids which originally focused on helium atoms,  were first addressed by Khalatnikov in 1957 using a hydrodynamic approach \cite{Kahl}.
Bassichis \cite{Bassi} and Nepomnyashchii \cite{YNep} employed the Bogoliubov method to study the superfluidity in Bose-Bose mixtures at zero temperature.
Later on, Colson and Fetter \cite{CFetter} generalized the Bogoliubov method to finite temperature.
Superfluidity in Bose-Fermi mixtures has been investigated in Refs (see e.g. \cite{Edw, Igor}).
Recently, observation of spin superfluidity in a Bose gas mixture has been reported in Ref \cite{Fava}.
The Andreev-Bashkin effect play a more prominent role in such mixture superfluids in both three (3D) and low dimensions \cite{Nesp}.
In 2D case, this effect may modify the usual Berezinskii-Kosterlitz-Thouless transition,
results in the occurrence of the vortex unbinding transition in one of the components can induce the breakdown of superfluidity in the other \cite{Karl}.
The two-fluid theory has been generalized for a superfluid in spin-orbit coupled Bose-Einstein condensate (BEC) with anisotropic effective masses \cite{Zhan}.

On the other hand, dipolar Bose-Bose mixtures has attracted tremendous attention due to the
intriguing role played by their anisotropic nature and long-range character. 
More recently, Er-Dy mixture has been experimentally achieved in two-species magneto-optical trap \cite{Ilz}.
Among the theoretical activity in binary mixtures with dipole-dipole interactions (DDI), one can quote:
rotonization \cite{Wilson}, solitons \cite{Adhik1},  supersolidity (see e.g. \cite{Wilson1, Yong}), quantum fluctuations \cite{Past, Boudj000},
miscible-immiscible phase transition \cite{Kumar}, and dipolar mixture droplets \cite{Boudj000}.

The purpose of the present paper is therefore to investigate the properties of dipolar Bose mixture superfluids.  
We use the hydrodynamic approach which has been successfully applied to study the properties of both 
one component BEC (see for book \cite{PethSmith, PitaevString, Boudjbook, ZNG}) and coupled BEC \cite{Past, Nesp, Zhan}.
At zero temperature, the hydrodynamic equations can be derived from the standard Gross-Pitaevskii equation
and thus, the hydrodynamic approach is valid only in the long-wavelength excitations regime. 
At finite temperatures, the validity of the theory requires fast thermalization times due to collisional relaxation \cite{ZNG, Boudj2016, Bouch}.
One of most important feature of the hydrodynamic theory is that it can correctly describe the infrared divergence of the longitudinal susceptibility in the low-energy and low-momentum limit, 
and the Nepomnyashchii-Nepomnyashchii identity \cite{YNep, Past, Wat}. 
These conditions cannot be satisfied in the usual Bogoliubov theory, and the many-body $T$-matrix theory, as well as the random-phase approximation\cite{Wat}.
 
Within the hydrodynamic theory we derive useful formulas for the excitations energy, the superfluid fraction, the static structure factor, 
and the first-order correlation function (one-body density matrix) in both dipolar and nondipolar mixtures. 
We show that the one-body density matrix is tending to a constant at large distances signaling the existence of the long-range order and thus, the formation
of  true condensates.  The role of the interspecies DDI and temperature in the one-body density matrix is discussed. 
Furthermore, it is found that the intra- and interspecies DDI may crucially reduce the superfluid fraction of the whole mixture and render it anisotropic.
The first transport measurements on an anisotropic superfluid in a single dipolar BEC has been recently reported in \cite{Wen}.
For a highly imbalanced mixture,  the superfluid behavior disappears in the dilute component.
Whereas,  in the case of a balanced mixture,  the superfluid fraction becomes identical to that of the one-component Bose liquid.
In quasi-2D case, we derive useful dispersion relations which exhibit the roton-maxon structure. 
Effect of the rotonization induced by the interspecies dipolar interactions on the static structure factor and the superfluid density is highlighted.

The rest of the paper is organized as follows. In Sec.\ref {Hyd}, we introduce the main features of the hydrodynamic approach
for a dipolar Bose-Bose mixture. We derive general formulas for the excitations spectrum which we employ to  
analyze the coherence and  the superfluidity.
As an application of our formalism, we demonstrate some concrete examples namely for 3D homogeneous case (Sec.\ref{3D}), 
and quasi-2D homogeneous mixture (Sec.\ref{2D}).
Finally, we summarize our results in Sec.\ref{concl}.



\section{Hydrodynamic Approach} \label{Hyd}

Consider two-component dipolar Bose-Bose mixture  labeled ($j=1,2$) in the weakly interacting regime. 
The dynamics of such a system is governed by the following nonlocal coupled GP equations
\begin{align} \label{GPE}   
i\hbar \dot{\Phi}_j ({\bf r},t) &=  h_j^{sp} \Phi_j ({\bf r},t)+\int d{\bf r'} V_j({\bf r}-{\bf r'}) n_j ({\bf r'},t) \Phi_j({\bf r},t) \nonumber \\
&+\int d{\bf r'} V_{12} ({\bf r}-{\bf r'}) n_{3-j} ({\bf r'},t) \Phi_j ({\bf r},t),  
\end{align}
where $h_j^{sp}=-(\displaystyle\hbar^2/\displaystyle 2m_j) \Delta + U_j({\bf r})$ with $m_j$ being the atomic mass
and $U_j({\bf r})$  external traps,  $V_j ({\bf r})$ and $V_{12}({\bf r})$ are respectively, the intraspecies and interspecies two-body interactions potentials. 
The gas density is defined as $n({\bf r})=|\Phi({\bf r})|^2$, and $\Phi({\bf r})$ is the condensate wavefunction.

Let us now introduce the density-phase formalism which requires to write the condensate wavefunction for each component in the form: 
\begin{equation} \label {eq9}
\Phi_j( \mathbf r,t)=\sqrt{n_j (\mathbf r,t)} \exp [i \phi_j (\mathbf r,t)], 
\end{equation}
where $\phi_j$ encodes the phase which is real and related to the superfluid velocity as $v_j=(\hbar/m_j) \nabla \phi_j$. 
After having substituting (\ref {eq9}) in the coupled nonlocal GP equations (\ref{GPE})  and separating real and imaginary parts, one obtains 
the continuity and the Euler-like equations, respectively
\begin{equation}  \label{Hyd1} 
\frac{\partial n_j} {\partial t} +{\bf \nabla}\cdot (n_j v_j)=0, 
\end{equation}  
and 
\begin{align}  \label{Hyd2} 
m_j\frac{\partial v_j}{\partial t} &=-{\bf \nabla} \bigg [-\frac{\hbar^2}{2m_j} \frac{\Delta \sqrt{n_j}} {\sqrt{n_j}} +\frac{1}{2} m_j v_j^2 +  U_j({\bf r})\\
&+ \int d{\bf r'} V_j({\bf r}-{\bf r'}) n_j ({\bf r'}) \nonumber\\
&+\int d{\bf r'} V_{12} ({\bf r}-{\bf r'}) n_{3-j} ({\bf r'}) \bigg], \nonumber
\end{align}
where $\Delta \sqrt{n_j}/ \sqrt{n_j}$ represents the so-called quantum pressures.

We consider small fluctuations of the condensed density $\hat n_j=n_j (\mathbf r)+\delta \hat n_j$
where  $\delta \hat n_j/n_j\ll1$. We then linearize Eqs.(\ref{Hyd1}) and (\ref{Hyd2}) 
with respect to $\delta \hat n_j$ and  ${\bf \nabla} \hat \phi_j$ around the stationary solution. \\
The zero-order terms yield:
\begin{align}  
\mu_j&=-\frac{\hbar^2}{2m_j} \frac{\Delta \sqrt{n_j}} {\sqrt{n_j}} + U_j({\bf r})
+ \int d{\bf r'} V_j({\bf r}-{\bf r'}) n_j ({\bf r'}) \nonumber\\
&+\int d{\bf r'} V_{12} ({\bf r}-{\bf r'}) n_{3-j} ({\bf r'}), 
\end{align}
where $\mu_j$ is the chemical potential for each component. \\
The first-order terms provide equations for the density and phase fluctuations:
\begin{align}  
&\hbar \frac{\partial} {\partial t} \frac{\delta \hat n_j} {\sqrt{n_j}} =\bigg[{\cal L}_j+ \int d{\bf r'} V_j({\bf r}-{\bf r'}) n_j ({\bf r'}) \label{F:td1}  \\
&+\int d{\bf r'} V_{12} ({\bf r}-{\bf r'}) n_{3-j} ({\bf r'}) \bigg] 2\sqrt{n_j} \,\hat \phi_j,  \nonumber\\ 
&2\sqrt{n_j}\,\hbar \frac{\partial \hat \phi_j} {\partial t} =-\bigg[{\cal L}_j+ \int d{\bf r'} V_j({\bf r}-{\bf r'}) n_j ({\bf r'}) \label{F:td2}  \\
&+\int d{\bf r'} V_{12} ({\bf r}-{\bf r'}) n_{3-j} ({\bf r'}) \bigg] \frac{\delta \hat n_j} {\sqrt{n_j}} \nonumber \\
&- 2 \int d{\bf r'} V_j({\bf r}-{\bf r'}) \sqrt{n_j ({\bf r})}  \, \delta \hat n_j ({\bf r'}) \nonumber \\
&- 2 \int d{\bf r'} V_{12} ({\bf r}-{\bf r'}) \sqrt{n_j ({\bf r})} \, \delta \hat n_{3-j} ({\bf r'}), \nonumber
\end{align}
where ${\cal L}_j=-(\hbar^2/2m_j) \Delta + U_j({\bf r}) -\mu_j$.

\subsection{Elementary excitations} 

Writing the phase and the density in the basis of the excitations
\begin{align}  \label{phs}
\hat \phi_j({\bf r})=(-i/2\sqrt{n_j ({\bf r})})  \sum_k [f_{jk}^{+}({\bf r})\, \exp (-i \varepsilon_k t/\hbar ) \, \hat {b}_{jk} - {\text H.c.}]
\end{align}
and 
\begin{align}  \label{den}
\delta \hat n_j({\bf r})=\sqrt{n_j ({\bf r})} \sum_k [f_{jk}^{-}({\bf r}) \exp (-i \varepsilon_k t/\hbar ) \, \hat {b}_{jk}+ {\text H.c.}],
\end{align}
where $\int d {\bf r} [f_{jk}^{+} {f_{jk'}^{-}}^*+ f_{jk}^{-} {f_{jk'}^{+}}^*]=2\delta_{k,k'}$.
Inserting Eqs.(\ref{phs}) and (\ref{den}) into (\ref{F:td1}) and (\ref{F:td2}), we then obtain the nonlocal Bogoliubov-de Gennes equations (BdGE):
\begin{align}  
\varepsilon_k f_{jk}^{-} ({\bf r})&=\bigg [{\cal L}_j  + \int d{\bf r'} V_j({\bf r}-{\bf r'}) n_j ({\bf r'})   \label{B1:td}  \\ 
&+\int d{\bf r'} V_{12} ({\bf r}-{\bf r'}) n_{3-j} ({\bf r'})  \bigg] f_{jk}^{+} ({\bf r}), \nonumber
\end{align}
and 
\begin{align}  
\varepsilon_k f_{jk}^{+} ({\bf r})& =\bigg [{\cal L}_j  + \int d{\bf r'} V_j({\bf r}-{\bf r'}) n_j ({\bf r'})   \label{B2:td}  \\ 
&+\int d{\bf r'} V_{12} ({\bf r}-{\bf r'}) n_{3-j} ({\bf r'})  \bigg] f_{jk}^{-} ({\bf r})  \nonumber \\
&+ 2\int d {\bf r'} \sqrt{n_j ({\bf r'})}  V_j({\bf r}-{\bf r'}) \sqrt{n_j ({\bf r})}  f_{jk}^{-} ({\bf r'}) \nonumber \\
&+ 2\int d {\bf r'} \sqrt{n_{3-j} ({\bf r'})}  V_{12} ({\bf r}-{\bf r'}) \sqrt{n_j({\bf r})}  f_{jk}^{-} ({\bf r'}). \nonumber
\end{align}
Equations (\ref{B1:td}) and (\ref{B2:td}) form a complete set to calculate the ground state and collective modes of dipolar Bose binary mixtures.
They reduce to the usual BdG equations for vanishing inter- and intra-species DDI.

From now on, we consider a homogeneous mixture with equal masses $m_1=m_2=m$.

Therefore, the functions $f_j$ take the form $f_{jk}^{\pm}=\left (\varepsilon_{1,2 k}/E_k \right)^{\pm 1/2}$ where
$E_k=\hbar^2k^2/2m$, and the spectrum corresponding to the BdG equations (\ref{B1:td}) and (\ref{B2:td}) 
is found to be composed of upper ($\varepsilon_{1k}$) and lower ($\varepsilon_{2k}$) branches 
\begin{equation} \label {Bog}
\varepsilon_{\pm k}= \sqrt{E_k^2+2E_k \mu_{\pm}  ({\bf k})}, 
\end{equation}
where
$\mu_{\pm } ({\bf k})=  \tilde V_1({\bf k}) n_1\, F_{\pm}({\bf k})/2$, 
$F_{\pm}({\bf k}) = 1 + \alpha \pm \sqrt{ (1-\alpha)^2 +4 \Delta ^{-1}\alpha }$,\,  $\alpha({\bf k})=   V_2({\bf k}) n_2/ V_1 ({\bf k}) n_1$, and 
\begin{align} \label {mc}
\Delta({\bf k}) &=\frac{\tilde V_1({\bf k}) \tilde V_2({\bf k})} { \tilde V_{12}^2({\bf k})},
\end{align} 
is the miscibility parameter, 
where $\tilde V_{ji} ({\bf k})$ is the Fourier transform of the two-body interaction potential.
For $\Delta ({\bf k}) >1$, the mixture is miscible while it is immiscible for $\Delta ({\bf k})<1$.
The spectrum (\ref{Bog}) is identical to that found in our recent paper \cite{Boudj00} using the Hartree-Fock-Bogoliubov  approximation.

\subsection{Superfluidity} \label{Supfluid}

The expression of the superfluid fraction can be given  in the frame of the hydrodynamic approximation \cite{Boudj4,Boudj5}.
Assuming that the two condensates are now moving with the same particle flow velocity ${\bf v}= \hbar {\bf p}/m$ \cite {YNep, CFetter}.
As in the case of a single dipolar Bose gas, the total momentum in each component is represented by a tensor as \cite{Nik,Boudj4,Boudj5, BoudjDp}
$Pj=- {\bf v_j} \int  ( d N_{jk}/d \varepsilon_{\pm k}) \, {\bf p_j} \otimes {\bf p_j}  \,  d {\bf p_j}/ (2 \pi \hbar)^3$, 
where $N_{jk}=[\exp(\varepsilon_{jk}/T)-1]^{-1}$ are occupation numbers for the excitations.  
The normal fraction of the dipolar Bose mixture liquid turns out to be given by 
\begin{equation}\label {supflui}
\frac{n_{nj}}{n_j} = \int  \frac{d \bf k}{ (2\pi)^3}  \frac{\hbar^2}{ n_j T m }  \frac{k \otimes k}{4 \sinh^2 (\varepsilon_{ \pm k}/2T)}.
\end{equation}
This clearly indicates the superfluidity  depends on the direction of the superfluid motion with respect to the orientation of the dipoles. 
In a single component dipolar BEC, the anisotropic behavior of the superfluidity can be observed by moving an attractive laser beam through the condensate \cite{Wen}.
In the absence of the DDI and interspecies interaction, the expression (\ref{supflui}) reduces to that derived earlier in \cite{LL9, Boudj3}.
It is worth noticing  that the zero-temperature spectrum (\ref{Bog}) remains valid for the calculation of the finite-temperature superfluid density \cite {CFetter}.

\subsection{First-order correlation function}

At equal times, the first-order correlation function is defined as
\begin{equation}\label {1Corr1}
g_j^{(1)} ({\bf s}, 0)=n_j+\langle \hat{\bar\psi}_j^\dagger({\bf r},0) \hat{\bar\psi}_j ({\bf r'},0) \rangle,
\end{equation}
where $s=|{\bf r-r'}|$.
For the second term in the right hand side of this equation, one has to use the transformation
\begin{align}\label {psibar}
\hat{\bar\psi}_j ({\bf r})&= \sum_k \bigg [\bigg(\frac{f_{jk}^{+}({\bf r}) + f_{jk}^{-} ({\bf r}) } {2}\bigg)  \hat b_{jk} e^{i {\bf k}.{\bf r}-i\varepsilon_{\pm k} t /\hbar} \\
&-\bigg(\frac{f_{jk}^{+}({\bf r}) - f_{jk}^{-}({\bf r})}{2}\bigg) \hat b_{jk}^\dagger e^{-i {\bf k}.{\bf r}+i\varepsilon_{\pm k} t/\hbar} \bigg], \nonumber
\end{align}
where $\hat b^\dagger$, $\hat b$ are operators of excitations. Employing the expectation values $\langle \hat b^\dagger \hat b \rangle=N_k$ and 
$\langle \hat b \hat b \rangle=\langle \hat b^\dagger \hat b^\dagger \rangle=0$, the $d$-dimensional one-body density matrix takes the form
\begin{align}\label {1Corr2}
g_j^{(1)} ({\bf s}, 0)&=n_j+ \int_0^{\infty}  \frac{d^d k}{(2\pi)^d}\bigg[ \frac{1}{4}\bigg(\frac{\varepsilon_{\pm k}} {E_k}+\frac{E_k}{\varepsilon_{\pm k}}-2\bigg) \\
&+ \frac{1}{2}\bigg(\frac{\varepsilon_{\pm k}} {E_k}+\frac{E_k}{\varepsilon_{\pm k}}\bigg)  N_{jk}\bigg]  \nonumber
e^{i {\bf k \cdot s}}.
\end{align}
The one-body density matrix is important for describing the coherence of the mixture.

\subsection {Static structure factor}

The static structure factor  of the mixture is connected to the density fluctuations 
$S_j({\bf k})= \langle \delta \hat n_j \delta \hat  n_j\rangle/n_j$ \cite{PitaevString}. 
According to Eq.(\ref{den}), we can write $S_j({\bf k})$ in momentum space as:
\begin{equation}\label {SFac}
S_j({\bf k})= \frac{E_k}{\varepsilon_{\pm k}} \coth \left(\frac{\varepsilon_{\pm k}}{2T}\right).
\end{equation} 
At zero temperature, Eq.(\ref{SFac}) reduces to $ S_j({\bf k})= E_k/\varepsilon_{\pm k}$.
At $n_j g_j \gg T$, we have the ideal gas result for the structure factor except in the phonon regime ($k\rightarrow 0$) where $S_j(k) \simeq T/\mu_{\pm}(|{\bf k} |=0)$ \cite{Boudj2}.
In the opposite situation, at $n_j g_j \ll T$, $ S_j({\bf k})$ simplifies to its zero temperature value except in the limit $k\rightarrow 0$ where 
the structure factor approaches its asymptotic value \cite{PitaevString}.
At large momenta, $ S_j({\bf k})$ approaches unity \cite{Boudj2}.
The static structure factor gives insight into the quantum and thermal fluctuations which are of pivotal importance in lower dimensions.

\section{Three-dimensional case} \label{3D}

As a concrete illustration, we apply our hydrodynamic formalism to the 3D uniform Bose mixture.
In 3D geometry,  the two-body interactions potential  reads
\begin{equation}  \label{InterPot}
V_{ji}({\bf r})=g_{ji}\delta({\bf r})+d_j d_i \frac{1-3\cos^2\theta} { r^3},
\end{equation}
where $g_{ji}=g_{ij}= 4\pi \hbar^2a_{ji}/ m_j $ correspond to the interspecies 
short-range part of the interaction, they characterized by the interspecies $s$-wave scattering lengths $a_{ji}=a_{ij}$,
$d_j$ stands for the magnitude of the dipole moment of component $j$, and
$\theta$ is the angle between the polarization axis-$z$ and the relative separation of the two dipoles.
In the Fourier space, the interaction potential (\ref{InterPot}) can be written as $\tilde V_{ij}(\mathbf k)=g_{ij}^{3D} [1+\epsilon_{ij}^{dd} (3\cos^2\theta_k-1)]$  \cite{Boudj000}, 
where $\epsilon_{ji}^{dd}=r_{*ji}/3a_{ji}$ with $r_{*ji}= m_j d_j d_i /\hbar^2$ being the characteristic dipole-dipole distance. 
Therefore, the above parameters turn out to be given as
$$\alpha({\bf k})=  \beta \frac{1+\epsilon_2^{dd} (3\cos^2\theta-1)}{1+\epsilon_1^{dd} (3\cos^2\theta-1)},$$
and
\begin{align} \label {mc}
\Delta({\bf k}) &=\frac{\tilde V_1({\bf k}) \tilde V_2({\bf k})} { \tilde V_{12}^2({\bf k})} \\
& =\Delta \frac{ [1+\epsilon_1^{dd} (3\cos^2\theta-1)] [1+\epsilon_2^{dd} (3\cos^2\theta-1)] } {[1+\epsilon_{12}^{dd} (3\cos^2\theta-1)]^2}, \nonumber
\end{align} 
where $\beta= n_2 g_2 /n_1 g_1$ and $\Delta=g_1 g_2/g_{12}^2$ is the miscibility parameter of a nondipolar mixture.
In the dipole dominated regime $\epsilon_j^{dd} >1$,  a uniform dipolar Bose mixture is unstable to collapse, 
whereas for $0<\epsilon_j^{dd} <1$, it is stable.

At low momenta ($k \rightarrow 0$), the Bogoliubov excitations behaving as sound waves and thus, the total dispersion relation reads 
\begin{equation} \label{sound}
\varepsilon_{\pm k}= \hbar c_{\pm} (\theta) k,
\end{equation} 
where the sound velocities $c_{\pm}$  are
 \begin{equation} \label{sound1}
c_{\pm} ^2 (\theta)=\frac{1}{2} \left[ c_1^2+c_2^2 \pm \sqrt{ \left( c_1^2-c_2^2\right) ^2 + 4 \Delta^{-1} c_1^2 c_2^2} \right] (\theta),
\end{equation}
where $c_j (\theta)= \sqrt{\tilde V_j(|{\bf k} |=0)  n_j /m}$ is the sound velocity of a single condensate. 
We see that the sound velocity (\ref{sound1}) is angular dependence due to the anisotropy of the DDI \cite{Boudj000}. 

For a nondipolar mixture ($\epsilon_j^{dd}=\epsilon_{ji}^{dd}=0$) and at zero temperature, the one-body density matrix (\ref{1Corr2}) turns out to be given as 
\begin{equation}\label {1Corr3}
g_j^{(1)} (s)=n_j+\frac{1}{ \sqrt{\pi}} n_1 \sqrt{n_1 a_1^3} \left(\frac{F_{+}}{ 2} \right)^{3/2} \left(\frac{\xi_{+}}{s}\right)^2; \,\,\,\,\, s \rightarrow \infty.
\end{equation}
where $\xi_{+}=\hbar/\sqrt{m \mu_{+}(|{\bf k} |=0)}$. For $g_{12}=0$, the function (\ref{1Corr3}) reduces to that of a single BEC.
It is important to see that $g_j^{(1)}$ tending to a constant value at large distances ($s \rightarrow \infty$),
indicating the existence of the long-range order in the upper branch of the mixture.

At very low temperatures, one can put $\coth(y) =1/y$. Then, at small momenta $k\rightarrow 0$, we obtain for the first-order correlation function
\begin{equation}\label {1Corr4}
g_j^{(1)} (s)=n_j+\frac{\Lambda^2}{2s};\,\,\,\,\,\,\;\;\;\ s \rightarrow \infty,
\end{equation}
where $\Lambda=\sqrt{2\pi \hbar^2/mT}$ is the thermal de Broglie wavelength.
We see that the thermal contribution decays as $1/s$ at $s \rightarrow \infty$ and it exceeds the zero
temperature contribution to $g_j^{(1)} (s)$ since this latter decays as $1/s^2$. 
Furthermore, Eq.(\ref{1Corr4}) shows clearly the presence of the long-range order even at finite temperatures.

\begin{figure}
\includegraphics[scale=0.45]{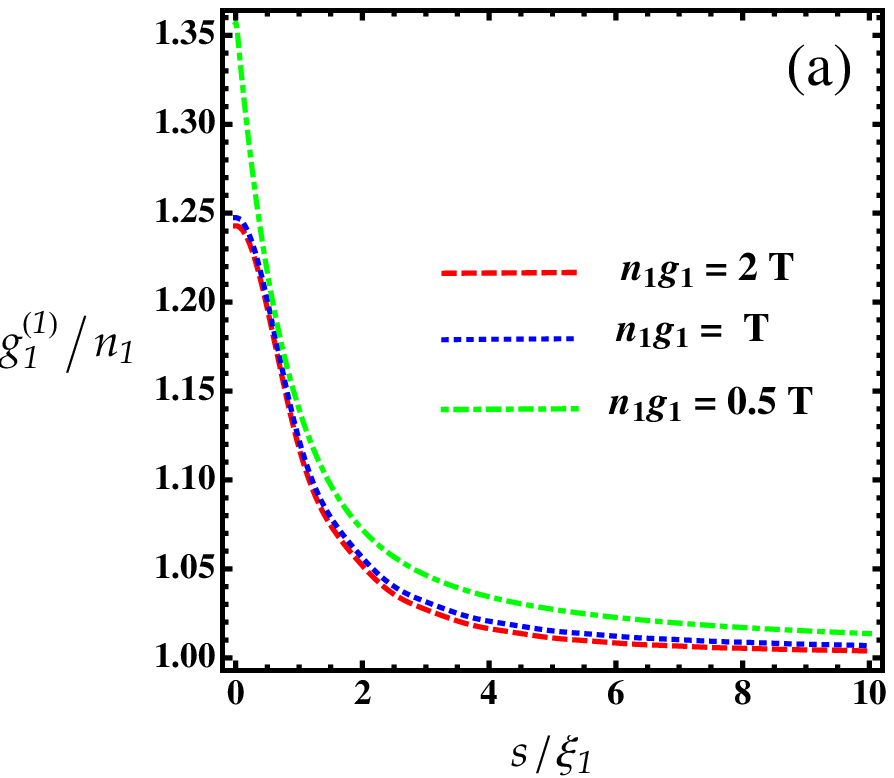}
\includegraphics[scale=0.45]{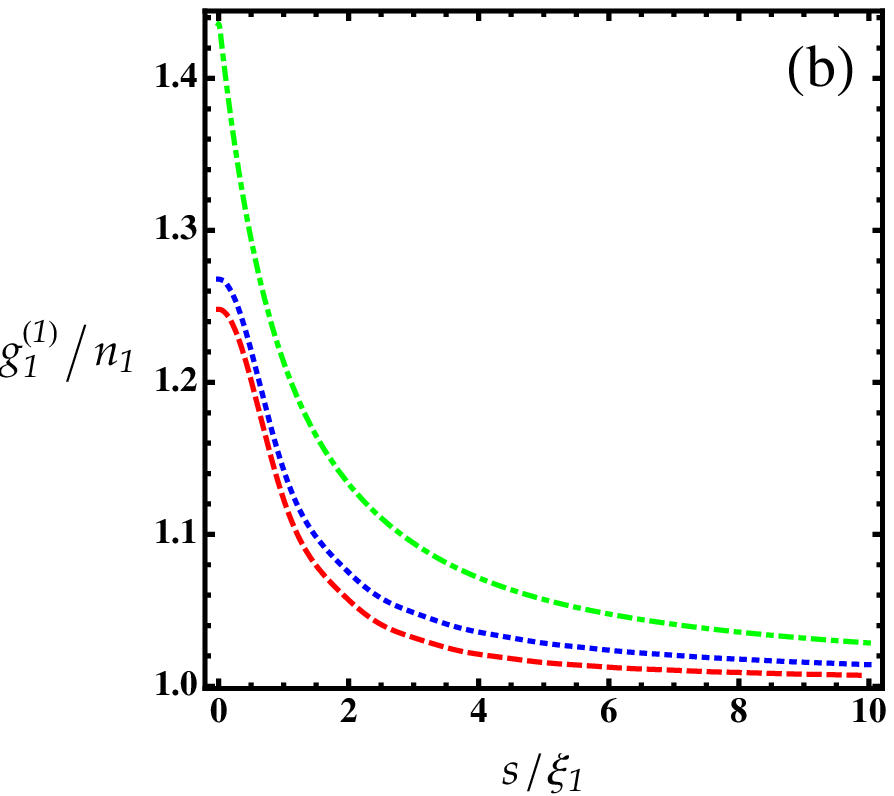}
 \caption{ (Color online) First-order correlation function associated with the upper branch for 3D mixture at different temperatures.
Parameters are: $r_{1*}=r_{2*}=131 a_0$, $n_2=10^{20}$ m$^{-3}$ \cite{Wen}, $a_1=a_2=141 a_0$ \cite {Tang}, 
$a_{12}=115.1 a_0$ ($a_0$ being the Bohr radius), $n_1=5\times 10^{20}$ m$^{-3}$, and (a) $\epsilon_{12}^{dd}=0.5$, (b) $\epsilon_{12}^{dd}=0.85$.
The interspecies scattering lengths $a_{12}$ and $\epsilon_{12}^{dd}$ can be varied by means of a Feshbach resonance.}
\label{cf} 
\end{figure}

For a dipolar mixture, the behavior of the one-body density matrix is also a decaying function at $s \rightarrow \infty$ and diverging near the center 
at relatively low temperatures $n_1g_1/T<1$ as is seen in Fig.\ref{cf} (a) showing the existence of the long-range order (i.e. existence of the two condensates). 
We observe also from the same figure that $g_1^{(1)} (s)$ is decreasing with increasing temperature.
The interspecies dipolar interactions $\epsilon_{12}^{dd}$ may enhance the first-order correlation function (see Fig.\ref{cf} (b)). 

As we have stated above, the tensorial superfluid fraction splits into a parallel and a perpendicular components. \\
In the parallel direction, expression (\ref{supflui}) gives for the superfluid fraction
\begin{align} \label{sup1}
 \frac{ n_{sj}^{\parallel}} {n_j}&= 1- \frac{4\hbar^2}{ n_j T m } \int \frac{d k\, d\theta}{(2\pi)^2}   \frac{k^4 \sin \theta \cos^2\theta}{4 \sinh^2 (\varepsilon_{\pm k}/2T)}, \nonumber\\
&=1- \frac{2\pi^2T^4}{45 m n_j\hbar^3}   \left[ \frac{{\cal J}_1^{-5 \parallel} (\epsilon_{dd})  }{c^5_+} + \frac{{\cal J}_2^{-5\parallel} (\epsilon_{dd})  }{c^5_-} \right].
\end{align}
In the perpendicular direction, the superfluid fraction can be computed from (\ref{supflui}) as:
\begin{align} \label{sup2}
 \frac{ n_{sj}^{\perp}} {n_j}&= 1-  \frac{4\hbar^2}{ n_j T m } \int \frac{d k \,d\theta}{ 8\pi^2} \frac{k^4 \sin \theta \sin^2\theta}{4 \sinh^2 (\varepsilon_{\pm k}/2T)} \nonumber\\
&=1- \frac{\pi^2T^4}{45 m n_j\hbar^3}   \left[ \frac{{\cal J}_1^{-5 \perp} (\epsilon_{dd})  }{c^5_+} + \frac{{\cal J}_2^{-5 \perp} (\epsilon_{dd})  }{c^5_-} \right].
\end{align} 
Note that expressions (\ref{sup1}) and (\ref{sup2}) are calculated at low temperature where the main contribution to the above integrals 
comes from the phonon branch. 
They show that the normal part of the mixture superfluid increases with $T$ whatever 
the values of $\epsilon_j^{dd}$, $\beta$ and $ \sqrt{n_{c1} a_1}$. 
Close to the transition, it coincides with the noncondensed density of a noninteracting Bose gas.

\begin{figure}
\includegraphics[scale=0.45]{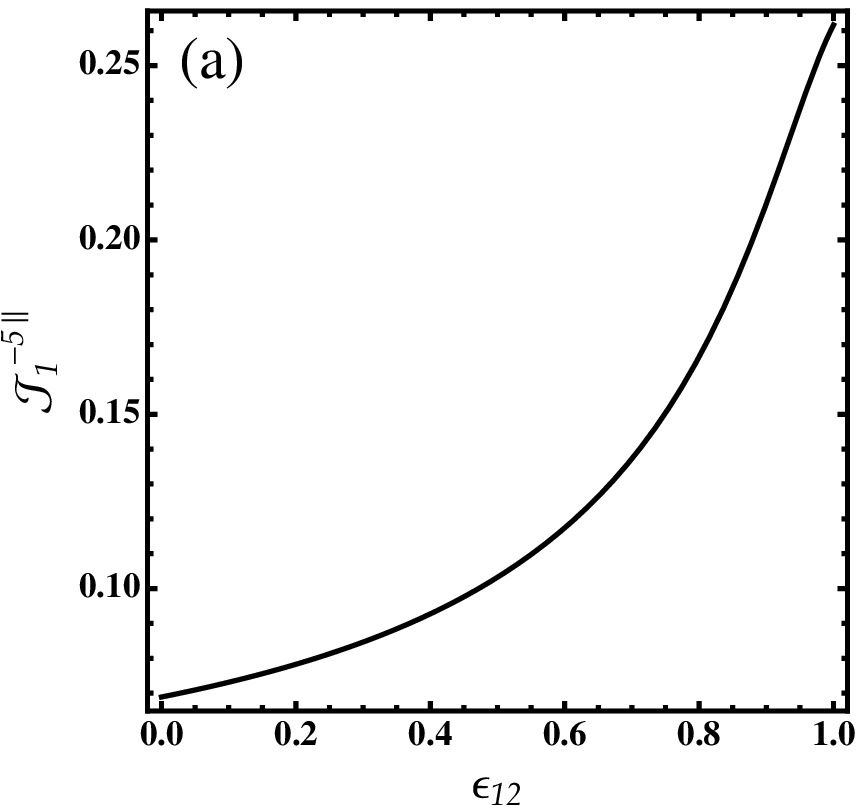}
\includegraphics[scale=0.44]{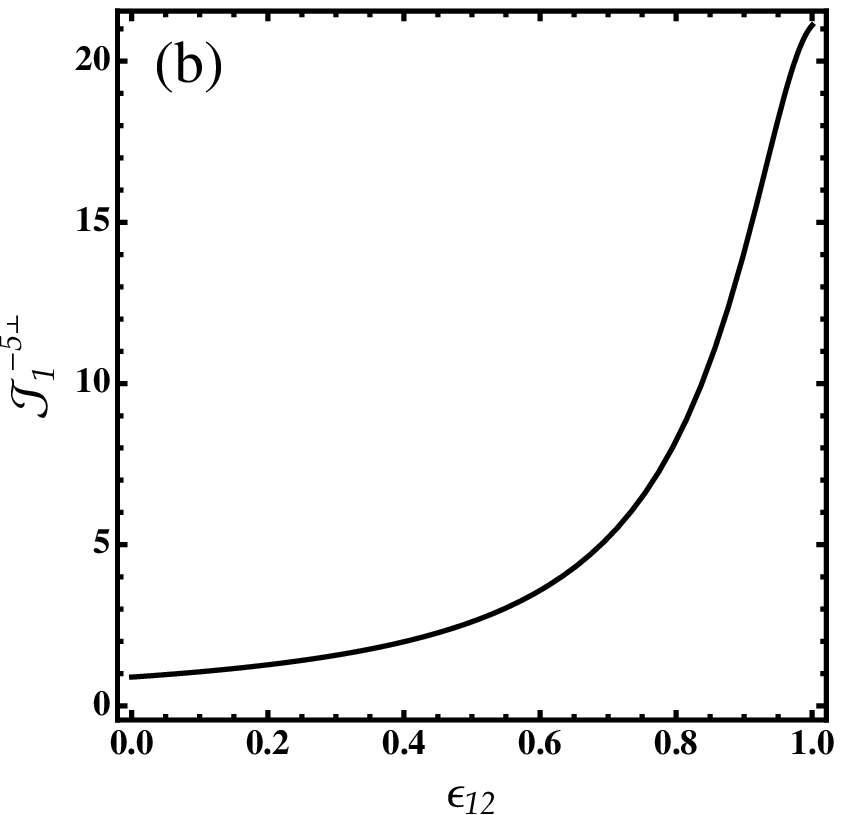}
\includegraphics[scale=0.45]{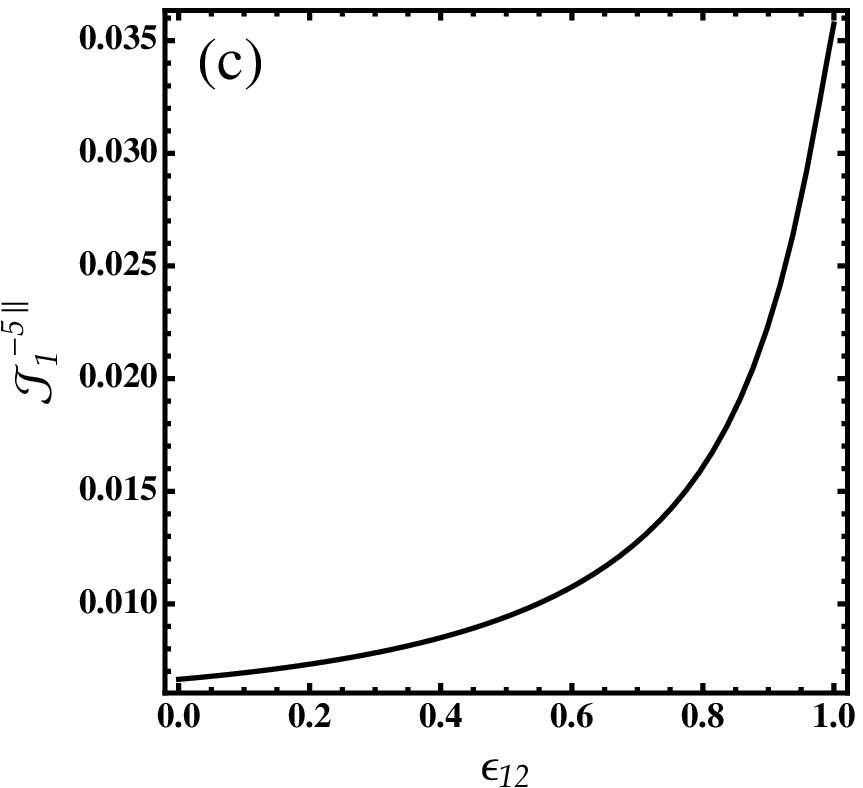}
\includegraphics[scale=0.44]{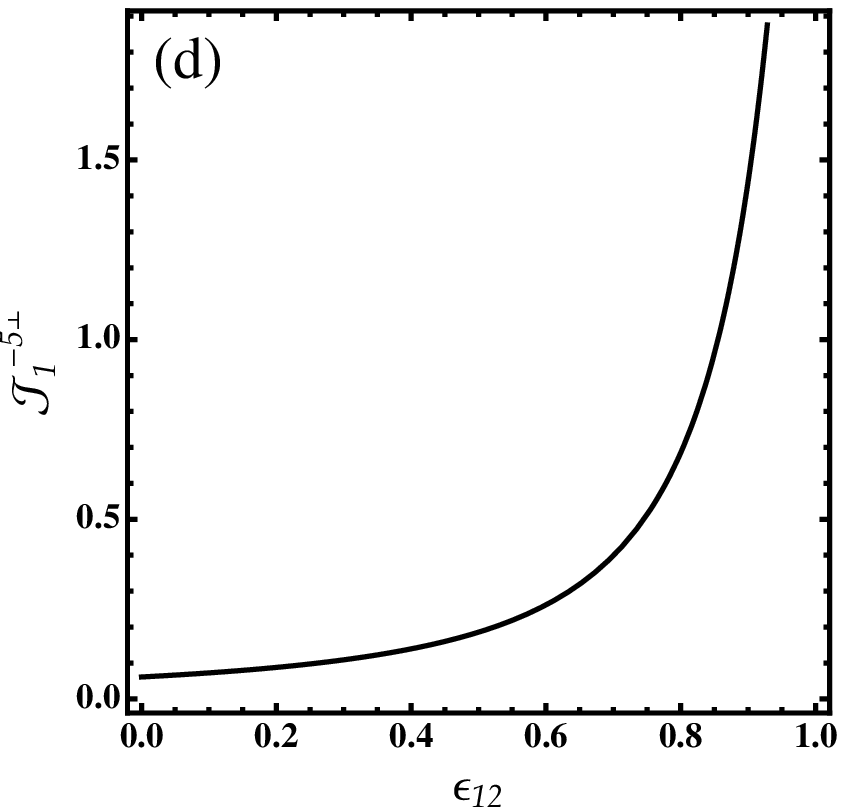}
 \caption{Functions ${\cal J}_1^{-5 \parallel} $, and ${\cal J}_1^{-5\perp} $ which describes the dependence of the
superfluid fractions associated with the upper branch  in the parallel  and perpendicular directions on the relative interspecies dipolar interaction strength $\epsilon_{12}^{dd}$.
(a)-(b)  $n_1=\times 10^{20}$ m$^{-3}$. (c)-(d) $n_1=5\times 10^{19}$ m$^{-3}$. Parameters are the same as in Fig.\ref{cf}. }
\label{spfd} 
\end{figure}

The functions ${\cal J}_j^{\ell} (\epsilon_{dd}) $ are defined as 
\begin{subequations}\label {Nfuncs}
\begin{align} 
{\cal J}_j^{\ell \parallel} (\epsilon_{dd})&= \int_0^{\pi}  \sin \theta  \cos^2 \theta \left[1+\epsilon_1^{dd} (3\cos^2\theta-1) \right]^{\ell/2} F_j^{\ell/2} \, d \theta, \\
{\cal J}_j^{\ell \perp} (\epsilon_{dd})&= \int_0^{\pi}  \sin^3 \theta \left[1+\epsilon_1^{dd} (3\cos^2\theta-1) \right]^{\ell/2}  F_j^{\ell/2} \, d \theta,
\end{align}
\end{subequations}
their behavior is displayed in Fig.\ref{spfd}.

Figures \ref{spfd} (a) and (b) show that the interspecies DDI may strongly reduce the superfluid fraction associated with the upper branch in both directions.
We observe also that $n_s^{\perp}$ is smaller than $n_s^{\parallel}$ for any value of $\epsilon_{12}^{dd} $
since  ${\cal J}_1^{\ell \perp} (\epsilon_{dd}) \gg {\cal J}_1^{\ell \parallel} (\epsilon_{dd})$.
For large $\beta$, i.e. when the first component $(n_1)$ is dilute, the superfluid fraction rises in both directions and the whole mixture liquid 
becomes practically superfluid specifically in the parallel direction (see Figs.\ref{spfd} (c) and (d)). 
One can infer that the superfluidity occurs in the dilute component $(n_1)$ at ultralow temperatures.
A similar behavior arises in ${}^4$He-${}^6$He mixture superfluid \cite{CFetter,YNep}. 
One the other hand, the dipolar functions ${\cal J}_2^{\ell} (\epsilon_{dd})$ are divergent signaling the disappearance of the superfluidity 
in the component related to the lower branch which in agreement with the results of Ref.\cite{Past}.

For a balanced mixture where $n_1 = n_2=n$ and $\tilde V_1({\bf k}) =\tilde V_2({\bf k})= \tilde V_{12}  ({\bf k})= \tilde V  ({\bf k})$, one has 
$F_1=4$ and $F_2=0$, thus,  the spectrum of the upper branch is similar to the spectrum of a single component dipolar BEC, 
$\varepsilon_{k}= \sqrt{E_k^2+8 E_k n \tilde V({\bf k})}$, while the spectrum associated with the lower branch becomes identical to that of free particles, $\varepsilon_{2k}=E_k$.
In such a case the superfluid fractions in the parallel direction reduce to 
\begin{equation}\label {supflui1}
\frac{n^{\parallel}_s}{n} =1-  \frac{\sqrt{2}}{8}\frac{2\pi^2T^4}{45 m n\hbar^3 c^5} {\cal Q}_{-5}^{\parallel}  (\epsilon_{dd})+ \frac{\Lambda^3 \zeta(3/2)}{n},
\end{equation}
where ${\cal Q}_{\ell}^{\parallel} (x)= \int_0^1 dy y^2 (1-x+3xy^2)^{\ell/2}$, 
have the properties ${\cal Q}_{\ell}^{\parallel} (0)=1/3$ and become imaginary for $x>1$.\\
In the perpendicular direction
\begin{equation}\label {supflui2}
\frac{n^{\perp}_s}{n} =1-  \frac{\sqrt{2}}{8}\frac{\pi^2T^4}{45 m n\hbar^3 c^5} {\cal Q}^{\perp}_{-5} (\epsilon_{dd})+ \frac{\Lambda^3 \zeta(3/2)}{n},
\end{equation}
where ${\cal Q}^{\perp}_{\ell}(x)=\int_0^1 dy (1-y^2) (1-x+3xy^2)^{\ell/2}$. 
Importantly, the superfluid fraction of a balanced mixture in both parallel and perpendicular directions is larger than that of one Bose liquid.
This can be attributed to the competition between the intra- and inter-component interactions.
We see also from Eqs.(\ref{supflui1}) and (\ref{supflui2}) that the normal part of the superfluid mixture in both directions
coincides  with the normal density of a dipolar one-component Bose fluid obtained at higher temperatures 
and with the noncondensed density of an ideal gas.
The difference between both quantities might be found by calculating higher order fluctuations corrections.

Near the phase separation  and at low temperature, the lower branch has the free-particle dispersion law $\varepsilon_{k-} = E_k$ \cite {CFetter}  
while the upper branch is phonon-like $\varepsilon_{k+} =\hbar c_1 (1 + \alpha)^{1/2} k$. 
This results indicates that the total normal density in the parallel direction has two different temperature dependence form: $n_n^{ \parallel}=p^{\parallel} T^4+ q\,T^{3/2}$, where 
$p^{\parallel}= \left(\sqrt{2} \pi^2/360 m n\hbar^3 c^5 \right) \int _0^{\pi}  d \theta \sin \theta  \cos^2 \theta / (1+\alpha)^{5/2}$,
and $q= (m /2\pi \hbar^2)^{3/2} \zeta (3/2)$. Whereas, in the perpendicular direction: $n_n^{\perp}=p^{\perp} T^4+ q\,T^{3/2}$, where 
$p^{\perp}= \left(\sqrt{2} \pi^2/360 m n\hbar^3 c^5 \right) \int _0^{\pi}  d \theta \sin ^3\theta / (1+\alpha)^{5/2}$.
This obviously indicates that the component related to the lower branch is extremely dilute.

\section{Quasi-two-dimensional case} \label{2D}

We focus our attention now on the homogeneous quasi-2D Bose-Bose mixture.
It is well known that in such systems, thermal fluctuations can destroy the long-range order associated with BEC at low temperatures,
but cannot suppress superfluidity in an interacting system  \cite{Boudj3, BoudjG, merm, pop, hoh, GPS}.

Let us consider a dipolar bosonic mixture confined in quasi-2D geometry by means of an external harmonic potential in the direction perpendicular to the motion (pancake trap).
All dipoles are assumed to be aligned perpendicularly to the plane of their translational motion throuhg through a strong electric (or magnetic) field. 
In momentum space, the two-body interaction potential reads \cite{BoudjG,Boudj6}
\begin{equation}\label {2DInt}
\tilde V_{ji}(\mathbf k)= g_{ji}^{2D}  (1- C_{ji}^{dd} \, k),
\end{equation}  
where the 2D short-range coupling constant is $g_{ji}^{2D} =g_{ji} /\sqrt{2} l_0$ with $l_0=\sqrt{\hbar/m \omega}$, 
$\omega$ is the confinement frequency, and $C_{ji}^{dd} =2\pi d_j d_i/g_{ji}^{2D}$ 
are dimensionless relative strengths describing the interplay between the DDI and short-range interactions.

The miscibility parameter becomes
\begin{align} \label {mc2D} 
\Delta({\bf k})=\Delta^{2D} \frac{ (1- C_1^{dd} \, k) (1- C_2^{dd} \, k)} {(1- C_{12}^{dd} \, k)^2}, 
\end{align} 
where  $\Delta^{2D}=g_1^{2D} g_2^{2D}/(g_{12}^{2D})^2$ is the miscibility parameter for 2D nondipolar mixture.
One should stress that in addition to the usual (long-wavelength) immiscibility-miscibility transition, the system is also sensitive to a supplementary transition originates
due to the roton modes \cite{Wilson}. In our case, such a transition could be controlled by the parameter $C_{ji}^{dd}$.

\begin{figure}
\centerline{
\includegraphics[scale=0.8]{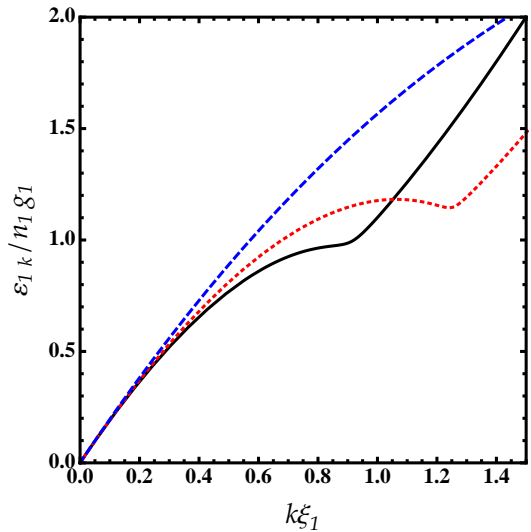}}
 \caption{ (Color online) Upper branch excitations energy of quasi-2D Bose mixture as a function of momentum $k$. 
Parameters are : $C_1^{dd}/\xi_1 =C_2^{dd}/\xi_1 = 0.5$, $n_2=10^{13}$ m$^{-2}$, $n_1=5\times 10^{13}$ m$^{-2}$ \cite{BoudjG}.  
Solid line: $C_{12}^{dd}/\xi_1 = 0.8$. Blue dashed lines: $C_{12}^{dd}/\xi_1 = 0.5$. Red dotted line: $C_{12}^{dd}/\xi_1 = 0.2$. 
Here $\xi_1=\hbar/\sqrt{mn_1g_1}$ is the standard healing length.}
\label{spec} 
\end{figure}

The behavior of the Bogoliubov dispersion relation is presented in Fig.\ref{spec}.
The upper branch of the two-component dispersion (\ref{Bog}) has a roton-maxon structure. 
The height of the roton depends on the interspecies DDI, $C_{12}^{dd}$, as is seen in the same figure. 
It worth noticing  that in regime of the low momenta, the spectrum is sound wave, where the sound velocity  is isotropic in a stark contrast to the 3D case. 
The excitations energy of the lower branch is unstable.

The correlation function can be calculated from  Eq.(\ref {1Corr2}).
Again, for a mixture with pure contact interactions and at $T=0$, one finds
\begin{equation}\label {1Corr5}
g_j^{(1)} (s)=n_j+\frac{m g_1^{2D} n_1} { 2\pi \hbar^2}  \frac{F_{+}}{ 2}  I_1 (s/\xi_{+}) K_1 (s/\xi_{+}),
\end{equation}
where $I_1$ and $K_1$ are the modified Bessel functions. 
In the absence of the inter-component interaction, the correlation function simplifies to that of a single 2D BEC.
At distances, $s \gg \xi_{+}$, Eq.(\ref{1Corr5}) yields for the correlation function $g_j^{(1)} (s)=n_j \left[ 1+ (m g_1^{2D} n_1/ 8\pi \hbar^2) F_{+}  (\xi_{+}/s) \right] $
signaling that at $T=0$, the long-range order still survive in the upper branch of quasi-2D Bose mixtures. 

\begin{figure}
\centerline{
\includegraphics[scale=0.8]{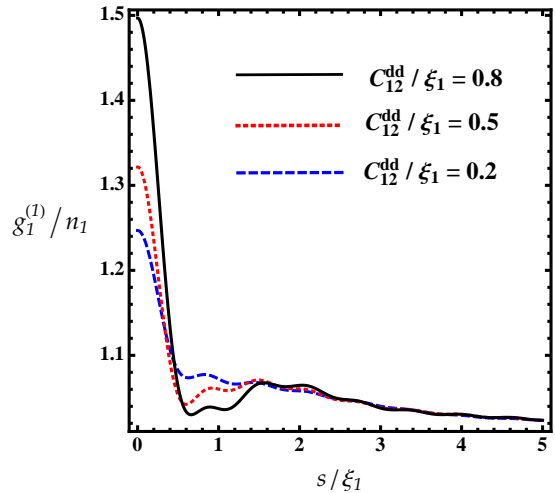}}
 \caption{ (Color online) First-order correlation function associated with the upper branch for 2D dipolar Bose-Bose mixture for different values of $C_{12}^{dd}/\xi_1$.
Parameters are the same as in Fig.\ref{spec}.}
\label{cf1} 
\end{figure}

For a dipolar mixture, the numerical simulation of the one-body density matrix shows that at $T=0$,  $g_j^{(1)} (s)$
decays and tends to its asymtotic value $n$ at large distances similarly to the nondipolar case (see  Fig.\ref {cf1}) 
showing the existence of the long-range order results in the formation of  quasi-2D two-component BEC and superfluidity.
By increasing $C_{12}^{dd}$, and close to the roton region,  $g_j^{(1)} (s)$ has an oscillatory behavior at large distances. 
This is a signature to the transition to a novel state of matter \cite{BoudjG}.
Note that, at finite temperature, the first-order correlation function of both dipolar and nondipolar mixtures undergoes a slow power law decay at $s \rightarrow \infty$.
Therefore, the long-range order is destroyed by the phase fluctuations prohibiting the formation a true BEC in 2D geometry.

\begin{figure}
\centerline{
\includegraphics[scale=0.8]{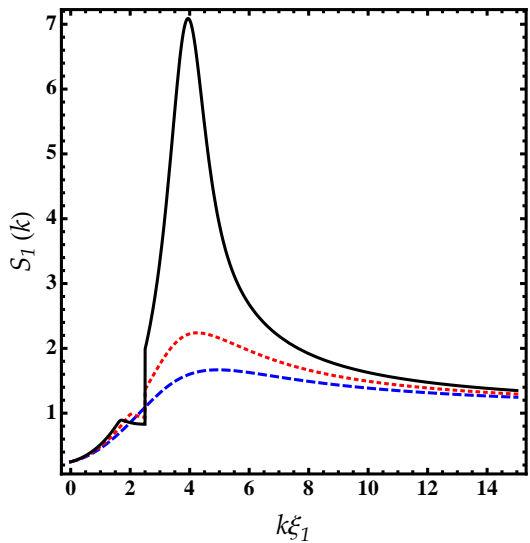}}
 \caption{ (Color online) Static structure factor from Eq.(\ref{SFac}) as a function of dimensionless variable $k\xi_1$ for different values of $C_{12}^{dd}/\xi_1$
 at $T/n_1 g =0.5$. Parameters are the same as in Fig.\ref{spec}.}
\label{SF2D} 
\end{figure}

As one can see in Fig \ref{SF2D}, as the interspecies interactions is increased,  $S({\bf k})$ develops a peak in the region of $k  \simeq 4/\xi$
indicating that the condensates start to disappear. 
By further increasing $C_{12}^{dd}$, the quantum fluctuations get more and more stronger and the peak becomes sharper which may lead 
to destroy the off-diagonal long-range order.
The temperature can also modify the height of the peak of structure factor.


The superfluid fraction can be obtained easily from Eq.(\ref{supflui})
\begin{equation}\label {supflui3}
\frac{n_{sj}}{n_j} =1-  \frac{3\zeta(3)}{2\pi \hbar^2 m n_j c_+^4} T^3,
\end{equation}
which is similar to that obtained for a single component BEC in Refs \cite{Boudj3, FHoh}.
Equation (\ref{supflui3}) indicates that the interspecies interactions $C_{12}^{dd}$ may crucially increase the thermal fluctuations and hence,  
decrease the superfluid fraction even at temperatures below the Kosterlitz-Thouless transition temperature.
In the roton regime, one can expect that the superfluid density might be strongly depressed \cite{BoudjG,Boudj33}.

\section{Conclusion} \label{concl}

We theoretically investigated the superfluid properties of weakly interacting homogeneous dipolar binary Bose condensates.
Within the realm of the hydrodynamic approach, we derived useful expressions for the excitations spectrum, the first-order correlation function,
the static structure factor, and the superfluid density.
We examined  in addition the impact of  temperature and interspecies  DDI  on the correlation function and on the superfluid density in both 3D and 2D geometries.
We showed that the first-order correlation function decays at larger distances at any temperature in the 3D case indicating the existence of the long-range order. 
In 2D case, it goes to its asymptotic value at large distances. 
Our analysis revealed also that in the roton regime, the correlation function has an oscillating behavior at large distances signaling the destruction of the long-range order
associated with the BEC. 
The static structure factor which contains information on the density fluctuations is decreasing with the interspecies interactions and develops a peak 
at relatively low momenta near the roton minimum. 
Moreover, we found that $\epsilon_{dd}^{12}$ may crucially reduce the superfluid fraction of the whole mixture.
For a highly imbalanced mixture,  the superfluid behavior disappears in the dilute component.
Whereas, in the case of a balanced mixture, the superfluid fraction becomes identical to that of the one-component Bose liquid.
In 2D case, the normal density of the mixture liquide is enhanced owing to the strong fluctuations induced by the interspecies interactions.
Our findings could provide important inputs to first measurements of a dipolar mixture superfluid. 
Promising candidates for the creation of such ensembles are  ${}^{164}$Dy-${}^{162}$Dy \cite{Kumar} and ${}^{162}$Dy-${}^{164}$Er \cite{Ilz}
which have almost equal masses.

In closing this paper, we beg to introduce some remarks on the employed hydrodynamic approach.
First, the theory needs further investigations for the lower branch excitations. 
Second, the hydrodynamics of dipolar mixture collapses owing to phonon and roton instabilities still remain also challenging.

\section{Acknowledgements}
We thank Dmitry Petrov for insightful discussions.
We would like to thank the LPTMS, Paris-sud for a visit, during which part of this work has been conceived.

\end{document}